\documentclass{elsarticle}

\usepackage{hyperref}
\PassOptionsToPackage{version=3}{mhchem}
\usepackage{mhchem}
\sloppypar

\journal{Chemical Physics Letters}

\date{April 11, 2014}
\begin{document}

\begin{frontmatter}

\title{Comment on 'Enhancement in the production of nucleating clusters due to dimethylamine and large uncertainties in the thermochemistry of amine-enhanced nucleation' by Nadykto et al., Chem. Phys. Lett., doi:10.1016/j.cplett.2014.03.036 (2014)}

\author[mymainaddress]{Oona Kupiainen-M\"a\"att\"a\corref{mycorrespondingauthor}}
\ead{oona.kupiainen@helsinki.fi}
\author[mymainaddress]{Henning Henschel}
\author[mysecondaryaddress]{Theo Kurt\'en}
\author[mymainaddress]{Ville Loukonen}
\author[mymainaddress]{Tinja Olenius}
\author[mymainaddress]{Pauli Paasonen}
\author[mymainaddress]{Hanna Vehkam\"aki}
\address[mymainaddress]{University of Helsinki, Department of Physics, P.O. Box 64, FI-00014 University of Helsinki, Finland}
\address[mysecondaryaddress]{University of Helsinki, Department of Chemistry, P.O. Box 55, FI-00014 University of Helsinki, Finland}
\cortext[mycorrespondingauthor]{Corresponding author}


%
%

\begin{abstract}
We comment on a study by Nadykto et al. recently published in this journal. Earlier work from our group has been misrepresented in this study, and we feel that the claims made need to be amended. Also the analysis of Nadykto et al. concerning the implications of their own density functional calculations is incomplete. We present cluster formation simulations allowing more conclusions to be drawn from their data, and also compare them to recent experimental results not cited in their work.
\end{abstract}


\end{frontmatter}

\section{Introduction}

Nadykto et al. \cite{Nadykto14} present Gibbs free energies of formation from density functional calculations for hydrates of the \ce{H2SO4*DMA} and \ce{(H2SO4)2*(DMA)_{1-2}} clusters incorporating up to 5 water molecules. This is a useful addition to their earlier publication \cite{Nadykto11}, where they presented data for the first two hydrates of these clusters. They also seem to have re-evaluated some of the earlier published clusters, although this is not explicitly stated.

A major fraction of the article is devoted to a critique of an earlier study by Paasonen et al. \cite{Paasonen12} and of the quantum chemistry methods used by Paasonen et al. \cite{Paasonen12} and Loukonen et al. \cite{Loukonen10}. The conclusion of Nadykto et al. that different quantum chemical methods may give even qualitatively different Gibbs free energies of formation for some clusters is by no means new and surprising, and has in fact been discussed on several occasions both by Nadykto et al. \cite{Nadykto11} and our group \cite{Kurten11nadykto,Kupiainen13}. By now, a number of far more extensive comparisons involving a large number of different methods are available, which found both the PW91 method used by Nadykto et al. and the B3LYP//RI-CC2 method used by our group to be unsatisfactory \cite{Leverentz13,Bork14}. 

While the conclusion that different methods sometimes predict different results is entirely uncontroversial, Nadykto et al. have misrepresented and misquoted our work to an extent that we feel impelled to make a number of clarifications. Also, some conclusions drawn by Nadykto et al. seem unsupported and misleading, which is why we present results of cluster formation simulations using the quantum chemical data of Nadykto et al.

\section{Misrepresentation of earlier studies}

Nadykto et al. state that our group \cite{Paasonen12} has developed an ATHN (apparently standing for Amine Ternary Homogeneous Nucleation) theory. We certainly do not claim to have developed a new nucleation theory, but rather have presented a series of simulations probing how the particle formation rate depends on different factors such as base concentration, relative humidity (RH), temperature or the identity of the base (dimethylamine vs. trimethylamine).

According to Nadykto et al. we {\em conclude} that the clusters are not hydrated. On the contrary, we state clearly that most of the \ce{H2SO4*DMA} clusters are hydrated \cite{Paasonen12}. However, in some simulations we {\em assumed} that the hydration of larger clusters does not have a strong impact on particle formation rates, as hydration energies for those clusters were not available at the time. These have, however, been calculated later, and cluster formation simulations \cite{Almeida13} confirm that, according to our quantum chemical data, hydration does not have a strong effect on particle formation rates in the sulfuric acid/DMA system. A more detailed study of the computed hydrate distributions can be found in ref.  \cite{Henschel14}.

Several details related to the computational chemistry methods used by Paasonen et al. \cite{Paasonen12} and Loukonen et al. \cite{Loukonen10} are presented incorrectly. It should be noted that, despite the unfortunate wording of Ortega et al. \cite{Ortega12} where the term “locally developed method” originates from, the combination of geometry optimizations and frequency calculations at one level with single-point energy calculations at a higher level is a standard approach in computational chemistry, see for example \cite{Foresman96}. Neither of the individual methods used by us (B3LYP/CBSB7 for geometry optimizations and frequencies and RI-CC2/aug-cc-pV(T+d)Z for electronic energies) are “locally developed” -- the only choice we have made is to combine these two standard methods. Especially B3LYP/CBSB7 is extremely well established and tested, as it is the method used for geometry and frequency calculations in the common CBS-QB3 combination method.  The RI-CC2 method is admittedly less “conventional” for single-point energy calculations, as it is primarily developed for excitation energies -- we originally chose it because it represented the highest level of electron correlation at that time available in the cost-effective Turbomole program suite. However, as CC2 represents a (minor) improvement on the well-established MP2 method, the criticism of Nadykto et al. is excessive. Furthermore, in the study by Loukonen et al. \cite{Loukonen10}, the single-point energy is calculated using the RI-MP2 method, not the RI-CC2 method as claimed by Nadykto et al. Paasonen et al. \cite{Paasonen12} and later studies (e.g. \cite{Almeida13}) do not use scaling factors, unconventional or otherwise. 

The statement of ``variations in the stepwise hydration free energies [being] ~ 7-10
kcal mol$^{-1}$'' in ref.  \cite{Loukonen10} is misleading. Of the stepwise hydration free energies only two are between 7 and 8 kcal/mol but none are in the range 8--10 kcal/mol, and also the changes between consecutive stepwise hydration energies are mostly less than 7 kcal/mol.   The values allegedly taken from ref. \cite{Paasonen12} that are presented in Figure 6 of Nadykto et al. \cite{Nadykto14} deviate somewhat from the values published in the original study. They also seem to be incorrect in Figure 5 of Nadykto et al. \cite{Nadykto14}, which, however, remains unclear due to the form of their presentation.

\section{Incomplete analysis of the data}

In Figure 3 of their article, Nadykto et al. \cite{Nadykto14} examine the hydrated fraction of each cluster type as a function of relative humidity. As both \ce{H2SO4*DMA} and \ce{(H2SO4)2*DMA} do indeed get hydrated to a considerable extent at atmospheric conditions, they infer that the presence of water affects cluster formation. The analysis made in their paper is, however, insufficient to determine {\em how} hydration affects cluster and particle formation. 

\begin{figure}
 \centering
 \includegraphics[width=.7\textwidth]{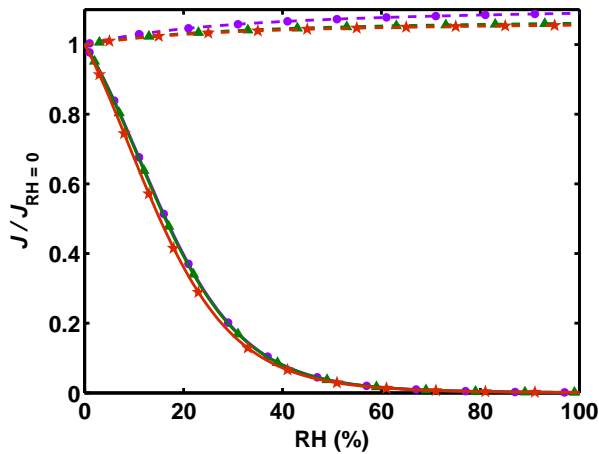}
 \caption{Simulated cluster formation rate $J_{1.3}$ as a function of RH, using the quantum chemical data of Nadykto et al. \cite{Nadykto14} (solid lines) and Almeida et al. \cite{Almeida13} (dashed lines), and the simulation methods described in ref. \cite{Almeida13}. The vapor concentrations are [\ce{H2SO4}] = $5\times10^6 \textrm{cm}^{-3}$ and [DMA] = 3 ppt (circles), 15 ppt (triangles) and 100 ppt (pentagrams).}
 \label{fig:JvsRH}
\end{figure}

Figure \ref{fig:JvsRH} of this Comment presents the RH dependence of the simulated formation rate of clusters containing at least 3 sulfuric acid and 2 DMA molecules using the thermochemical data of Nadykto et al. \cite{Nadykto14}. This formation rate is, from here on, denoted as $J_{1.3}$, as the \ce{(H2SO4)3*(DMA)2} cluster has a mobility diameter of about 1.3 nm. The simulation is similar to that presented in ref. \cite{Almeida13}, and a detailed explanation can be found there. The only difference is that since in the case presented here the growth proceeds in some conditions mostly by addition of single (possibly hydrated) acid and base molecules, the \ce{(H2SO4)3*(DMA)2} cluster is assumed to be the smallest stable cluster, while in the simulations based on our data in ref. \cite{Almeida13} the clusters grew mostly by addition of \ce{H2SO4*DMA*(H2O)_{1-5}} clusters and the criterion for particle formation was set to be the formation of \ce{(H2SO4)3*(DMA)3} clusters. As a comparison, we present also simulation results using our thermochemical data but the same formation criterion as for the data of Nadykto et al. 

From the simulations presented in Figure \ref{fig:JvsRH}, it can be concluded that according to the data of Nadykto et al. \cite{Nadykto14} hydration certainly has an effect on cluster formation -- it inhibits it effectively. This, perhaps surprising, behavior is explained by the fact that according to the quantum chemical data of Nadykto et al. \cite{Nadykto14}, while the \ce{(H2SO4)2*DMA} and, to some extent, \ce{(H2SO4)2*(DMA)2} clusters are hydrated, hydration does not stabilize them with respect to evaporation of hydrated sulfuric acid molecules and \ce{H2SO4*DMA} clusters, but rather increases their effective evaporation rate. Our thermochemical data also predicts an increase in effective evaporation rates of the \ce{(H2SO4)2*(DMA)_{1-2}} clusters due to hydration. However, even the hydrated clusters are predicted to be so stable that the increase in collision rates due to increased cluster size dominates the overall effect of hydration on the cluster formation rate $J_{1.3}$.

Nadykto et al. continue by analyzing the equilibrium concentrations of sulfuric acid---DMA clusters to assess the impact of DMA on sulfuric acid cluster formation. For the understanding of particle formation, this is, however, of limited value: if clusters are formed, they are not in equilibrium but  at most in a steady state. The equilibrium cluster distribution can be very different from the steady-state distribution relevant to the atmosphere and especially to new-particle formation. The only situation where small clusters may exist in something resembling an equilibrium distribution is when the vapor concentrations are so low that the formation of larger clusters is negligible, and also external losses are so low that their effect on the cluster distribution can be neglected.

Furthermore, instead of evaluating the {\em relative} concentrations of \ce{H2SO4*(DMA)_{0-2}} clusters with different DMA content, the {\em absolute} steady-state concentration of clusters containing two \ce{H2SO4} molecules and any number of DMA molecules at varying DMA vapor concentrations would be a more informative measure for the effect of DMA on cluster formation.

\section{Comparison to experiments}

Though lamenting the lack of relevant experimental studies, Nadykto et al. fail to mention the recent measurements of sulfuric acid---DMA clusters presented by Almeida et al. \cite{Almeida13}, where formation rates of 1.7 nm clusters ($J_{1.7}$) as well as concentrations of neutral clusters containing two \ce{H2SO4} molecules (and an unknown number of DMA and \ce{H2O} molecules) were measured at different sulfuric acid and DMA vapor concentrations. We present here a comparison of the thermochemical data by Nadykto et al. \cite{Nadykto14} with these experimental findings, and with our thermochemical data \cite{Almeida13}. It should be noted that the simulation results corresponding to our cluster energies differ slightly from those presented in ref. \cite{Almeida13}, since we use here the same set of clusters and the same formation criterion as for the simulations with the data of Nadykto et al. Specifically, we have now left out the clusters with three and four sulfuric acid molecules as well as all charged clusters, but have on the other hand included the hydrates, which were only used in one test simulation in ref. \cite{Almeida13}.

\begin{figure}
 \centering
 \includegraphics[width=.7\textwidth]{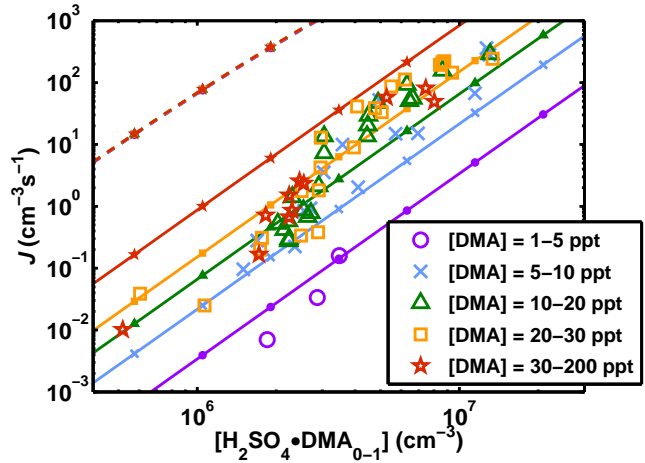}
 \caption{Comparison of measured formation rate of 1.7 nm clusters (big open symbols, \cite{Almeida13}) and simulated formation rates of 1.3 nm clusters (lines with small filled symbols) in conditions corresponding to the CLOUD chamber, using the quantum chemical data of Nadykto et al. \cite{Nadykto14} (solid lines) and Almeida et al. \cite{Almeida13} (dashed lines) and the ACDC dynamics code. The DMA concentrations of measurement points and simulations, respectively, are 1--5 or 3 ppt (circles), 5--10 or 8 ppt (crosses), 10--20 or 15 ppt (triangles), 20--30 or 25 ppt (squares) and 30--200 or 100 ppt (pentagrams). The relative humidity is 38\%.}
 \label{fig:JvsA}
\end{figure}

\begin{figure}
 \centering
 \includegraphics[width=.7\textwidth]{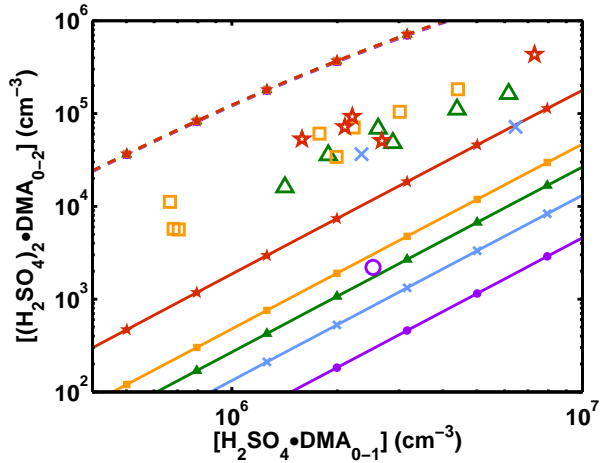}
 \caption{Comparison of measured (big open symbols, \cite{Almeida13}) and simulated (lines with small filled symbols) concentrations of clusters containing two \ce{H2SO4} molecules. See Figure \ref{fig:JvsA} for the explanation of the symbols.}
 \label{fig:dimers}
\end{figure}

Figure \ref{fig:JvsA} shows the measured formation rates $J_{1.7}$ of sulfuric acid---DMA clusters together with simulated formation rates $J_{1.3}$ using quantum chemical data from both Nadykto et al. \cite{Nadykto14} and Almeida et al. \cite{Almeida13}. The experimental formation rates were measured in the presence of ions produced by ambient background radiation at a rate of 4 ion pairs per second, but the effect of ions was concluded to be negligible based on the measurements. 

Some qualitative agreement can be seen between the measured formation rates and simulations based on the thermochemical data of Nadykto et al. \cite{Nadykto14}, although the DMA dependence seems to be overestimated by the simulation. The simulations based on our cluster energies, on the other hand, capture the weak DMA dependence but overestimate the value of the cluster formation rate.

However, the simulated formation rate corresponds to neutral clusters with a mobility diameter of approximately 1.3 nm, and therefore cannot be directly compared to the experimental rate. As clusters between 1.3 nm and 1.7 nm may evaporate back to smaller sizes or be lost by deposition on walls before reaching the diameter of 1.7 nm, the formation rate of 1.7 nm-diameter particles is likely to be somewhat lower than the formation rate of 1.3 nm particles.

A more direct comparison between experiments and simulations can be achieved by examining the concentration of clusters of the type \ce{H2SO4*(DMA)_{0-2}*(H2O)_{0-5}}. As DMA and water molecules evaporate during the detection of the clusters, only a sum over all clusters containing two \ce{H2SO4} molecules can be obtained from the measurements. 

Figure \ref{fig:dimers} presents a comparison of measured and simulated steady-state concentrations of these neutral two-acid clusters. Here the cluster formation energies presented by Nadykto et al. \cite{Nadykto14} fail to reproduce the experimental findings --- the simulated cluster concentrations are too low by up to a factor of approximately 100, suggesting that binding of the clusters is underestimated by the PW91 functional used by Nadykto et al.

\section{Conclusions}

We have here corrected a number of misrepresentations in the recent Letter by Nadykto et al. \cite{Nadykto14} criticizing our earlier work. We have also presented some cluster formation simulations using the thermochemical data published by Nadykto et al. and show that they predict a hindering effect of relative humidity on cluster formation. Furthermore, we have shown that although cluster formation rates predicted based on their data match reasonably well with experiments done at the CLOUD chamber, the stability of clusters containing two \ce{H2SO4} molecules is severely underestimated compared to experimental findings. We fully agree with Nadykto et al. that further experiments, especially reliable field measurements of amine concentrations, are needed to draw meaningful conclusions about the role of amines in atmospheric new-particle formation.

\section*{References}

\bibliography{../../artikkelit/oonan_viitteet}

\end{document}